\documentstyle[12pt]{article}
\renewcommand{\theequation}{\thesection.\arabic{equation}}
\newcommand{\nc}{\newcommand}
\nc{\beq}{\begin{equation}} \nc{\eeq}{\end{equation}}
\nc{\beqa}{\begin{eqnarray}} \nc{\eeqa}{\end{eqnarray}}
\nc{\lsim}{\begin{array}{c}\,\sim\vspace{-21pt}\\< \end{array}}
\nc{\gsim}{\begin{array}{c}\sim\vspace{-21pt}\\> \end{array}}

\nc{\scR}{{\cal R}}
\nc{\scL}{{\cal L}}
\nc{\al}{\alpha}
\nc{\ald}{\dot{\alpha}}
\nc{\be}{\beta}
\nc{\bed}{\dot{\beta}}
\nc{\lam}{\lambda}
\nc{\nud}{\dot{\nu}}
\nc{\lamd}{\dot{\lam}}

\newcommand{\drawsquare}[2]{\hbox{%
\rule{#2pt}{#1pt}\hskip-#2pt%  left vertical
\rule{#1pt}{#2pt}\hskip-#1pt%  lower horizontal
\rule[#1pt]{#1pt}{#2pt}}\rule[#1pt]{#2pt}{#2pt}\hskip-#2pt%  upper horizontal
\rule{#2pt}{#1pt}}% right vertical

\newcommand{\Yfund}{\raisebox{-.5pt}{\drawsquare{6.5}{0.4}}}%  fund
\nc{\vom}{ \left({v\over M}\right) }
\nc{\vomnb}{ {v\over M} }

\nc{\Ap}[2]{A^\prime_{#1#2}}

\nc{\Q}[2]{Q_{#1#2}}
\nc{\R}[2]{R_{#1#2}}
\nc{\Y}[2]{Y_{#1#2}}
\nc{\V}[2]{V_{#1#2}}
\nc{\q}[2]{q_{#1#2}}
\nc{\G}[2]{G_{#1#2}}
\nc{\W}[2]{W_{#1#2}}
\nc{\D}[2]{D_{#1#2}}
\nc{\A}[2]{A_{#1#2}}
\nc{\p}[2]{p_{#1#2}}
\nc{\vv}[2]{v_{#1}^{#2}}
\nc{\rr}[2]{r_{#1}^{#2}}
\nc{\lij}[2]{l_{#1}^{#2}}

\nc{\spa}{SU(2)_1}
\nc{\spb}{SU(2)_2}
\nc{\spc}{SP(2n-4)}
\nc{\spd}{SP(4n+2m-10)}
\nc{\Lsc}[2]{\scL_{#1#2}}
\nc{\Lrm}[2]{L_{#1#2}}
\nc{\Rsc}[2]{\scR_{#1#2}}

\begin{document}

\begin{titlepage}

\begin{center}

\vspace{2cm}

{\hbox to\hsize{hep-ph/9703312 \hfill  Fermilab-Pub-97/060-T}}

\bigskip

\vspace{2cm}

\bigskip

\bigskip

\bigskip

{\Large \bf
Gauge-Mediated Supersymmetry Breaking without 
Fundamental Singlets
}

\bigskip

\bigskip

 {\bf Yael Shadmi}  \\

\bigskip

\bigskip
\bigskip

{ \small 

\it Fermi National Accelerator Laboratory\\
  P.O.Box 500, Batavia\\
  IL 60510, USA\\

{\rm email}: yshadmi@fnal.gov \\
 }

\vspace{1.3cm}

{\bf Abstract}
\end{center}

\noindent
The messenger sector of existing models of 
gauge-mediated supersymmetry breaking may be simplified
by using a non-renormalizable superpotential term
to couple the vector-like quark and lepton 
messenger fields
to a chiral gauge-invariant of the supersymmetry-breaking sector.
This eliminates the need for a fundamental singlet
and for an additional gauge sector needed to generate appropriate
expectation values for the singlet component fields.
This scenario is more natural if the 
supersymmetry-breaking sector itself involves 
a non-renormalizable superpotential.
Several examples are constructed based on 
non-renormalizable 
$SU(n)\times SU(n-1)$
supersymmetry-breaking theories.

\end{titlepage}

\renewcommand{\thepage}{\arabic{page}}
\setcounter{page}{1}

\baselineskip=18pt

\section{Introduction}

Most models of gauge-mediated supersymmetry breaking rely on 
a singlet field $S$ with $A$- and $F$-type expectation values to
generate supersymmetry breaking masses for a pair of ``messenger
fields'', $f$ and $\bar f$, through the superpotential coupling
\beq
\label{s}
W = S \, f \cdot {\bar f}\ .
\eeq
With the fields $f$, $\bar f$ transforming as
a vector multiplet of the Standard Model (SM) gauge group, supersymmetry 
breaking is then communicated to the SM fields through the SM gauge 
interactions~\cite{dnns}.

It is usually non-trivial to generate appropriate expectation 
values for the singlet. To do that, the most economical models 
employ a $U(1)$ gauge symmetry sector with superpotential 
couplings to the singlet,
in addition to the basic supersymmetry-breaking 
sector~\cite{dnns}. 
Also, an $S^3$ term must be included in the superpotential to avoid 
runaway behavior.

But a generic supersymmetry-breaking theory contains different 
gauge-invariants with different $A$- and $F$-type vevs. 
It is therefore natural to try to use these to replace the fundamental
singlet.  
The field $S$ of eqn.~(\ref{s}) is then a composite, 
and the term~(\ref{s}) is a higher-dimension term, suppressed
by an appropriate power of some scale $M$. 
While the appearance of this scale is in general ad hoc,
some supersymmetry-breaking models inherently involve such a scale,
since they rely on non-renormalizable superpotentials to
achieve supersymmetry breaking~\cite{US},\cite{PST2},\cite{it}. 
Furthermore, as we will see below,
with a renormalizable theory as the supersymmetry-breaking sector,
these models are only viable when some dimensionless coupling
is taken to be extremely small, on the order of $10^{-9}$. 
This constraint can be alleviated if the supersymmetry-breaking sector
involves a non-renormalizable superpotential.
The reason for this the following.
Since the term~(\ref{s}) is suppressed by some power of the scale $M$,
and since we would like $M$ to be large,
the fields making up the composite singlet should have large 
expectation values
for the messenger mass scale to be of the correct order.
If the supersymmetry-breaking sector is non-renormalizable,
with terms suppressed by $M$,
the typical expectation values are naturally large.
However, in a renormalizable model, this requires some small coupling.
It should be stressed that all our examples do require some small coupling,
between $10^{-4}$ and $10^{-1}$ depending on the model we consider.

We present several examples 
of gauge-mediated supersymmetry-breaking models
in which the singlet field $S$ is replaced by a composite field 
of the supersymmetry-breaking sector.
As the supersymmetry-breaking sector we use  
a class of $SU(n)\times SU(n-1)$ gauge theories described in~\cite{US}.

There are several motivations for using these particular theories.
First, the $SU(n)\times SU(n-1)$ theories
involve non-renormalizable superpotentials
for $n > 4$, and therefore provide a natural setting for 
introducing the non-renormalizable term~(\ref{s})
as explained above.

Second, these theories have supersymmetry-breaking, 
calculable minima that may be studied through a simple sigma model.
In fact, it is possible to study many features of the minimum analytically,
and this will prove useful for the present analysis.

As an added bonus, the superpotentials of these theories do not 
conserve any $R$-symmetry. 
Hence, the models we construct are probably the only phenomenological 
examples with dynamical supersymmetry-breaking that do not resort to 
supergravity considerations in order to avoid a massless $R$-axion. 

A potentially  problematic feature of these models is 
that they contain massless fermions.
However, as we will see in section~\ref{specifics}, the massless
fermions do not pose any cosmological problem if the scale $M$ is
sufficiently big, as is the case in the examples we construct. 
It should be stressed that the existence of the massless fermions 
is not related to the focus of this paper,
namely, the possibility of eliminating the fundamental singlet. 
For example, we expect our qualitative results to hold for models based
on the analogous $SU(n)\times SU(n-2)$ supersymmetry-breaking theories
of~\cite{PST2}, which do not contain massless fermions.
In fact, our main results probably apply to a much larger class of theories,
since they follow from simple dimensional analysis.

We discuss the general requirements
on the models, and derive some general results based on 
dimensional analysis in section~\ref{general}. 
In section~\ref{specifics} we study some examples based on 
$SU(n)\times SU(n-1)$ supersymmetry-breaking theories.
Some technical details concerning the $SU(n)\times SU(n-1)$
minimum we consider are collected in the Appendix.

\section{Communicating supersymmetry breaking to the standard model}
\label{general}
\setcounter{equation}{0}

As outlined in the introduction, our models consist, apart from the
fields of the supersymmetric Standard Model,
of a supersymmetry-breaking sector (SB), and of the 
vector-like quark and lepton 
messenger 
fields~\cite{dnns} $f$ and ${\bar f}$, 
with the superpotential 
\beq
\label{fullw}
W\ =\ W_{SB} \ +\ S\, f\cdot {\bar f}\ .
\eeq
Here $W_{SB}$ is the superpotential of the supersymmetry-breaking model,
and
\beq
\label{sdef}
S\ =\ {1\over M^{d-1}} S^\prime \ ,
\eeq
where $S^\prime$ is a gauge-invariant combination of the fields 
of the supersymmetry-breaking sector, of dimension $d$.
The field $S$ is chosen so that it has both $A$- and $F$-type 
vacuum expectation values.
In the following, we will sometimes refer to these vevs as $S$ 
and $F_S$.

Let us now summarize the requirements on the expectation values
$S$ and $F_S$.

First, for the scalar messengers to have positive 
masses~\footnote{
Note that the messenger masses only depend on 
the absolute value of $F_S$,
but for  simplicity, we omit the absolute-value sign throughout this 
paper.}~\cite{dnns}, 
\beq
\label{fsoverssq}
F_S < S^2 \ .
\eeq
Second, to generate appropriate masses for the SM superpartners we 
need~\cite{dnns,martin}
\beq
\label{fsovers}
{F_S \over S} \sim 10^4 - 10^5~{\rm GeV} \ .
\eeq  
For brevity, we will require $F_S/S\sim 10^{4.5}~{\rm GeV}$.

Third, the most serious constraint on these models comes 
from the requirement that the supersymmetry-breaking scale 
is low enough. 
In principle, the   K\" ahler potential may contain higher dimension 
terms, suppressed by some power of $M$,
that couple either the standard-model fields, 
or the messenger fields, to the fields of the 
supersymmetry-breaking sector. 
Such terms could induce contributions of the order of
\beq
\label{m0}
m_0 \ = \ {F_0 \over M} \ ,
\eeq 
to the masses of the scalar messengers,
or to the masses of the SM scalar superpartners. 
Here $F_0$ is the supersymmetry-breaking scale squared.
As we will see shortly, 
when combined with~(\ref{fsoverssq}),
the requirement 
\beq
\label{m0one}
m_0 \ = \ {F_0 \over M} \sim 1~{\rm GeV}\ ,
\eeq
which would avoid problems with flavor-changing neutral currents,
can only be satisfied in the type of models we are
considering 
by taking one of the dimensionless couplings
that appear in the superpotential to be extremely small,
on the order of $10^{-9}$. 
Although not unnatural in the 
't~Hooft sense, 
since taking any of these couplings to zero typically restores 
some global symmetry, we find this unacceptably small.
Instead, we must assume that no higher-dimension terms 
that couple the SM fields and the supersymmetry-breaking sector fields
appear in the K\" ahler potential at the tree level. 
We will therefore take $M < M_{Planck}$.
Below we choose $M \leq M_{GUT}$.

The scenario we envision is that some new physics takes place above the 
scale $M$.
This new dynamics involves the fields of the supersymmetry-breaking model
(or just some of them) and the messengers,
and gives, as its low energy theory, 
the theory we describe with the superpotential~(\ref{fullw}).
It would of course be nice to have an actual microscopic theory that 
does this,
but at present we do not know of such an example.

While it is perhaps not unreasonable to assume that no terms coupling the 
SM fields
to the fields of the supersymmetry-breaking sector appear in 
the K\" ahler potential, 
one cannot assume the same for the messenger fields,
since these couple directly to the fields of the supersymmetry-breaking sector
through the superpotential.
It is therefore necessary to ensure that contributions to the messenger masses
from possible K\" ahler-potential terms, of the order $F_0/M$, are negligible 
compared to contributions coming from~(\ref{s}).
In fact, $F_0/M$ should be small compared to both the messenger masses
and their mass splittings,
in order to generate acceptable masses for the SM superpartners.
A non-zero value of ${\rm Str} M^2$, taken over the messengers,
may lead to negative masses squared for the SM squarks and 
sleptons, 
especially in models of the type we
are considering,
in which a large hierarchy of scales   
exists due to the presence of non-renormalizable 
terms suppressed by a large energy scale 
$M$~\cite{pt}, \cite{ptnew}, \cite{berkeley}, (see also~\cite{str}).
We therefore 
require\footnote{
The dangerous contribution to the supertrace is of the order
$\left({F_0\over M}\right)^2\,\log{\left({M^2\over S^2}\right)}$ 
and for $M=M_{GUT}$
the logarithm is approximately 5.
}$^,$
\footnote{
I thank S.~Trivedi for a discussion of this estimate.
}
\beq
\label{smallf0}
{F_0\over M} \leq 10^{-1} \, {F_S\over S} \ .
\eeq

Finally, one would like to have $\sqrt{F_0}\leq 10^{9}$~GeV,
so that supergravity contributions to the superpartner masses
are at most at the order of 1~GeV.
With~(\ref{fsovers}), (\ref{smallf0}), this is automatically satisfied for 
$M\leq 10^{15}$~GeV.
However, for $M = M_{GUT}$, the stricter bound,
\beq
\label{gutbound}
{F_0\over M} \leq 10^{-2} \, {F_S\over S} \,
\eeq
is needed, instead of~(\ref{smallf0}).

Let us now see what the  requirements
 (\ref{fsoverssq}), (\ref{fsovers}) and (\ref{smallf0})
imply for our models.
Here we will only present  rough order-of-magnitude
estimates. 
A more quantitative
analysis is undertaken in section~\ref{specifics} 
where specific examples are studied. 

Since the field $S$ is a composite field 
of dimension $d$, 
\beq
\label{svev}
S \sim  \ M \, \vom^d \ ,
\eeq 
where $v$ is the typical expectation value in the problem. 
We also have,
\beq
\label{fsvev}
{F_S\over S} \ \sim\   \vom^{-1}\, {F_0\over M} \ .
\eeq
If no large numerical factors appear in~(\ref{fsvev}),
eqn.~(\ref{smallf0}) (\ref{gutbound}), then imply
\beq
\label{voverm}
\vomnb \leq  10^{-1} \ or\  10^{-2}\ .
\eeq

Now let us assume that the 
highest-dimension term appearing in the superpotential
of the supersymmetry-breaking model, $W_{SB}$,
is also of dimension $d$.
In particular, for $d > 3$,
we assume that this highest-dimension term is necessary for
supersymmetry breaking to occur. 
Then the supersymmetry-breaking scale will typically be
of the order
\beq
\label{f0est}
F_0 \ \sim\ W/v 
\ \sim\ \al M^2\, {\vom}^{d-1}
\eeq
where $\al$ is the dimensionless coefficient of the 
highest-dimension term in the superpotential $W_{SB}$.
We then have
\beq
\label{fsest}
F_S \ \sim\ \alpha\, \ M^2\, \vom^{2\, (d-1)} \ ,
\eeq
and
\beq
\label{fsoversqscaling}
{F_S\over S^2} \ \sim\ \alpha \, \vom^{-2} \ .
\eeq
If no large numerical factors appear in~(\ref{fsoversqscaling}), 
we see from~(\ref{fsoverssq}) and~(\ref{voverm}), 
\beq
\label{smallcoupling}
\alpha \  \leq \ \vom^2 \ \leq 10^{-2}\ or   \ 10^{-4} \ .
\eeq

Thus, generically, some of the dimensionless couplings appearing in the 
supersymmetry-breaking superpotential $W_{SB}$ need to be small 
in order to satisfy both~(\ref{fsoverssq}), (\ref{smallf0}).

It is worth noting that, whereas the requirement~(\ref{voverm}) holds 
quite generally in the absence of large numerical factors,
(in fact, it is not much of a constraint, since $v$ should be much
smaller than $M$ for the analysis to be valid),
the condition~(\ref{smallcoupling}) depends sensitively on the assumption
that the highest dimension term in $W_{SB}$ is of the same dimension
as the composite $S$. 
In particular, if the dimension of the composite $S$
is smaller than the highest-dimension term in $W_{SB}$,
the condition~(\ref{smallcoupling}) may be avoided altogether.
However, 
as the examples we discuss in the next section demonstrate,
chiral gauge-invariant fields, or moduli,  
may scale in the same way with 
$v/ M$ even when they have different dimensions. 
The reason for this is simple--the different terms appearing 
in $W_{SB}$ are nothing but gauge-invariants, and at a generic minimum
these terms are comparable, so that the expectation values of 
the corresponding
gauge invariants only differ by  dimensionless couplings.

Finally, it would seem that~(\ref{smallcoupling}) may be avoided
if $F_S$ is suppressed compared to $S^2$. But that typically
means that $F_S$ is also suppressed relative to $F_0$, so that
the RHS of~(\ref{fsvev}) contains a small factor, which then enters 
squared in~(\ref{smallcoupling}), making matters worse.
One is therefore led to consider regions in which $F_S$ is not particularly
suppressed with respect to the other $F$ components in the problem.

At this stage, both $\al$ and $v/M$ are determined.
The messenger scale
\beq
\label{fsoversest}
{F_S\over S} \ \sim \
\al\, M\vom^{d-2} \ \sim\
 M\vom^d \ .
\eeq
 is then completely fixed in terms of the scale $M$.
Here we have used ~(\ref{svev}),
(\ref{fsest}), (\ref{smallcoupling}).
For example, for $M_{GUT}$, with~(\ref{gutbound}),
one needs $d= 6$ or 7 to obtain the desired messenger scale.
For $M=10^{15}$~GeV, with~(\ref{smallf0}),
one needs instead $d= 10$ or 11.
Thus, for these models to be viable, the supersymmetry-breaking
model must involve a non-renormalizable superpotential.

To summarize, the conditions~(\ref{fsoverssq}), (\ref{smallf0})
imply a specific relation between the coupling $\al$ and $v/M$
(see eqn.~(\ref{smallcoupling})).
Then, to generate the correct hierarchy between the messenger scale
and the scale $M$, it is necessary to have, for large $M$,
either a very small coupling, or a non-renormalizable
superpotential $W_{SB}$. 
Thus, by using a non-renormalizable supersymmetry-breaking sector,
one can avoid dimensionless couplings that are extremely small.
Indeed, for a renormalizable model, 
eqn.~(\ref{fsoversest}) gives, with $d=3$ and $M=M_{GUT}$,
$\al\sim 10^{-8}$.

In the next section 
we will therefore turn to specific examples with a 
non-renormalizable $SU(n)\times SU(n-1)$
supersymmetry-breaking sector.

\section {Models with an $SU(n)\times SU(n-1)$ supersymmetry-breaking sector}
\label{specifics}
\setcounter{equation}{0}

\subsection{The $SU(n)\times SU(n-1)$ theories}

As our supersymmetry-breaking sector we use the
 $SU(n)\times SU(n-1)$ gauge theories
of~\cite{US}.
These theories have the  matter content,
$Q \sim (\Yfund , \Yfund )$,
$L_I \sim (\overline{\Yfund}, {\bf 1})$, with $I=1\ldots n-1$ and 
$R_A \sim ({\bf 1}, \overline{\Yfund})$, with $A=1\ldots n$,
and the superpotential
\beq
\label{wtree}
W_{SB} = \lambda ~ \Sigma_{I} Y_{II} 
\ +\  \alpha~{b^1\over M^{n-4} }
\ +\   \beta~{b^n\over M^{n-4} } \ .
\eeq
where
$Y_{IA} = L_I \cdot Q \cdot R_A$, and
$b^A  = (R^{n-1})^A$ are the baryons of $SU(n-1)$.
(When appropriate,
all indices are contracted with  $\epsilon$-tensors).

In the presence of the superpotential~(\ref{wtree}), the original 
$SU(n-1)\times SU(n)\times U(1)\times U(1)_R$
global symmetry
is broken
to $SU(n-1)\times U(1)_R$
for $\lambda\neq 0$, which is further broken to 
$SU(n-2)\times U(1)_R$
for $\alpha\neq 0$. 
Finally, the last term in~(\ref{wtree}) breaks the $U(1)_R$ symmetry,
so that the remaining global symmetry is $SU(n-2)$.

As was shown in~\cite{US}, these theories 
break supersymmetry as long as $\al\neq 0$.
For $\al= 0$, the theories have runaway supersymmetric minima along 
the baryon flat directions,
and far  along these flat directions, 
the light degrees of freedom are weakly coupled~\cite{yuri}. 
Therefore, for large $M$,  the properties of the minimum can be reliably 
calculated~\cite{pt}.
In~\cite{pt}, this was used to study the minimum of the analogous
$SU(n)\times SU(n-2)$ theory (see also~\cite{berkeley} for the case of
$SU(n)\times SU(n-1)$).
We will therefore only outline the main points of the argument 
here, and refer the reader to~\cite{pt} for details.

Consider then a  $D$-flat direction with the 
fields $R_A$, with $A = 1\ldots n-1$, obtaining  expectation
values of order $v$. 
The gauge group $SU(n-1)$ is then completely broken at the scale $v$,
while the $SU(n)$ group remains unbroken. 
However, as a result of the first term in~(\ref{wtree}), 
all $SU(n)$ fields now get masses of order $\lambda v$.
For large enough $v$, these fields can be integrated out,
leaving, at low energies a pure $SU(n)$ which confines at the scale 
\beq
\label{lambdal}
\Lambda_L = \left ((\lambda v)^{n-1} \Lambda^{2n+1} \right)^{{1\over 3n}} 
\ .
\eeq
Below this scale, one is then left with the light components of the fields
$R$, with the $SU(n-1)$ dynamics negligibly
 weak, and the (strong) $SU(n)$ dynamics
decoupled, except that its non-perturbative contribution 
to the superpotential
\beq
\label{wnp}
\Lambda_L^3 = \left( \lambda^{n-1} b^n \Lambda^{2n+1} \right)^{{1\over n}} 
\ ,
\eeq  
arising from gaugino condensation in the pure $SU(n)$,
involves the fields $R$ (recall $b^n \sim (R_1\ldots R_{n-1})$).
As was argued in~\cite{yuri}, quantum corrections to the K\" ahler potential 
for the fields $R$ are very small, 
so that it is of the form
\beq
\label{kahlerr}
K\ = \ R^{\dagger A}\, R_A \ .
\eeq
Thus, all the properties of the vacuum may be calculated.

As in~\cite{pt}, we will find it convenient to work in terms of the baryons.
Our low energy theory  is then a theory of the  $n$ baryons $b^A$, 
with the superpotential 
\beq
\label{spotential}
W_{SB}\ = \
\left( \lambda^{n-1}\, \Lambda^{2n+1}\,  b^n \right)^{{1\over n}}
\ +\  \alpha~{b^1\over M^{n-4} }
\ +\   \beta~{b^n\over M^{n-4} } \ ,
\eeq 
and the K\" ahler potential
obtained from~(\ref{kahlerr}) as in~\cite{ads}, \cite{kahlerd}, 
\beq
\label{kahler}
K\ = \ (n-1)\,   (b^\dagger_A b^A)^{1\over n-1} \ .
\eeq

At the  minimum we consider,
the only baryons with non-zero vevs are $b^1$ and $b^n$.
It is convenient to 
define $r$ and $v$ such that 
\beq
\label{b1n}
b^1 = r \, b^n \ \ {\rm and}\ \ 
b^n = v^{n-1} \ ,
\eeq
where, as in the subsequent discussion, 
$b^A$ stands for the expectation value rather than the field.
The ratio $r$ is determined by the ratio
of dimensionless coupling  $\beta/\al$,
and is given in the appendix, where various details of the minimum 
are summarized.

We can then write the $F$-type expectation-values of $b^1$, $b^n$ as,
\beq
\label{barf}
F_{b^A} \ =\ F_A\,\alpha \, v^n \, \vom^{n-4} \ ,
\eeq
where
$F_{A=1,n}$, which are dimensionless functions of $n$ and $r$, 
are given in the appendix.

This is in fact all we need if we only wish to use the baryon operators
as our composite singlets.
It would also be useful however to consider the trilinears $Y_{IA}$ 
for this purpose.
Their vevs are given by (see appendix),
\beq
\label{yii}
Y_{II} \ = \ {n \, \al \over q \, \lam}\, M^3 \vom^{n-1} \ , \ \ \ 
Y_{1n} \ = \ r \, Y_{II}
\eeq
with $I=1\ldots n-1$, and where $q$ is a function of $n$ and $r$, 
given in the appendix.

Finally the $F$-type vevs of the fields $R$,
may be written as
\beq
\label{fr}
F_{R_A}\ = \   f_{r_A} \, \alpha \, M^2 \, \vom^{n-2} \ ,
\eeq
where again,  
$f_{r_A}$ are dimensionless functions of $n$ and $r$ and are 
given in the appendix.

\subsection{ $S =b^n/M^{n-2}$}

Choosing $S =b^n/M^{n-2}$ we have,
\beqa
\label{sbn}
 S \ &=&\ M\, \vom^{n-1} \ ,\\
\label{fsoversbn}
{F_S\over S} \ &=&\ F_n\, \al\, M\, \vom^{n-3} \ ,\\
\label{fsoverssqbn}
{F_S\over S^2} \ &=&\ F_n\, \al\,  \vom^{-2} \ .
\eeqa
To satisfy the requirements~(\ref{fsoverssq}) and~(\ref{smallf0})
without having very small couplings,
it is best to choose a region in which $F_S$ is not suppressed
compared to the other $F$ components in the problem, 
so that $F_n$ is order 1.
To see this, note that 
$${F_S\over S} \ \sim\ F_n\, {F_0 \over v} \ . $$
Therefore, the smaller the factor $F_n$ gets,
the smaller the value of $v$ that is needed to keep $F_0$ low.
Since $v$ enters squared in~(\ref{fsoverssqbn}), 
this would require a smaller coupling $\alpha$ as well.

We find that the optimal choice is  $r\sim 0.5$ 
(corresponding to $\beta/\al$ between 0.5 and 0.74 for $n=4\ldots 20$).
Taking $M=M_{GUT}$ and $n=8$, the different requirements on $F_S$ and $S$
can be met with $\al = 3.2\cdot 10^{-4}$ and  $v/M =2.4\cdot 10^{-2}$.
Alternatively, for $n=7$, one can take
$\al = 9\cdot 10^{-5}$, 
with $v/M = 1.3\cdot 10^{-2}$.

Note that since $M = M_{GUT}$, we use the stronger 
constraint~(\ref{gutbound}).
Taking instead $M=10^{15}$~GeV,
for which the less-stringent constraint~(\ref{smallf0}) can be used,
we find that for $n=12$ 
$\al = 7\cdot 10^{-3}$ and $v/M = 0.11$.
Raising $n$ to  $n=13$,
one can take
$\al = 1\cdot 10^{-2}$ 
with $v/M = 0.14$.
For all these choices, and in the following section,
$F_S/S = 10^{4.5}$~GeV, $F_S/S^2 = 0.75$
and $F_0 = 1-2\cdot 10^{18}$~GeV$^2$.

Choosing the baryon $b^1$, instead of $b^n$, to play the role
of the singlet leads to similar results.

It is amusing to note that these models contain gauge-invariant 
operators that are natural candidates for generating a $\mu$-term.
Consider for example the $SU(8)\times SU(7)$ model with $S = b^{8}/M^6$,
and add the superpotential term
\beq
\label{mu}
{1\over M^2} \, Y_{22}\, H_U\, H_D\ ,
\eeq
where $H_U$ and $H_D$ are the two Higgs doublets.
The $F$-vev of $Y_{22}$ vanishes for $r=0.57$. 
For this choice then, (\ref{mu}) generates a $\mu$-term but
no $B\mu$-term.
Also note,
\beq
\label{musize}
{Y_{22}\over S} = n q^{-1} \, {\al\over \lambda}\ ,
\eeq
so taking $\lam=1$, 
we get
$Y_{22} \sim 10^2$~GeV (where we also used the fact that $q^{-1}\sim 0.8$).

However, we have assumed throughout that the K\" ahler potential does not 
contain any terms that couple the SM fields to the fields of the 
supersymmetry-breaking
sector. 
Such terms, if present, would contribute masses of order $10^{2.5}$~GeV
to the SM scalars superpartners.
But this assumption would be quite implausible if we allowed superpotential
terms of the form~(\ref{mu}).

\subsection{ $S = Y/M^{2}$}

We can also take the trilinear invariants, $Y_{IA}$, 
to replace the singlet.
Here we take 
$S = Y_{1n}/M^{2}$, which turns out to be the optimal choice. 
For this choice we have 
\beq
\label{yvev}
S \ = \ {n \,  \over q \, \lam}\, M \vom^{n-1} \ ,
\eeq
where we used~(\ref{yii}), 
and,
%%%%%%%
\beq
\label{fsoversy}
{F_S\over S} \ =\ 
{ (1+r^2)^{{n-2\over 2(n-1)}}\over r}\,
f_{R_n}\,
M\, \vom^{n-3} \ .
\eeq
Here and throughout this section, we set the dimensionless baryon 
coupling, $\al$, to 1. 
As we will see, the ``small coupling'' in this case
is the Yukawa coupling $\lam$, multiplying the trilinear terms
in $W_{SB}$. 
Note that this coupling drops out in the ratio $F_S/S$,
but appears in the ratio  $F_S/S^2$.

Again, it is best to consider regions in which $f_{R_n}$ is not small,
and we choose $r=0.5$.
For $M=M_{GUT}$, one can take $n=10$, with 
$\lam = 4\cdot 10^{-3}$, 
and $v/M = 2.2\cdot 10^{-2}$.
Choosing instead $M=10^{15}$~GeV,
we need
$\lam = 1\cdot 10^{-2}$, 
$v/M = 8.8\cdot 10^{-2}$ with $n=13$,
and $\lam = 0.12$ 
with $v/M = 0.1$ for $n=14$.

Recall 
that to get the low-energy theory we are using, 
we have integrated out the fields $Q$ and $L$, 
assuming their masses,
$\lambda v$,
are much bigger than $\Lambda$, the scale of the $SU(n)$ group.
Since we are now considering small values of $\lam$,
we must make sure that the ratio (see appendix)
\beq
\label{scale}
{\Lambda\over \lambda v} \ =\
\left(n\, q^{-1} \al \lam^{-3} \vom^{n-4} \right)^{{n\over 2n+1}} \ 
\eeq
is still small.
It is easy to see that for sufficiently high values of $n$,
this is indeed the case.
Setting $\al =1$ and neglecting $q^{-1}$, which is order 1 for $r\sim 0.5$,
one can check that it is acceptably small in all our examples.

Note that the ``small coupling'' in this case is around 10$^{-3}$ 
for $M_{GUT}$,
and 10$^{-1}$ for $M^{15}$~GeV, an order of magnitude bigger than the 
 ``small coupling''  that is required when using the baryon as the singlet.
The difference is due to a numerical factor--essentially a factor of $n$
that enters the ratio $F_S/S^2$.

Finally, we note that for $M < 10^{15}$~GeV,
the typical size of the ``small coupling'' remains the same
(see eqn.~(\ref{smallcoupling})), 
but the value of $n$ goes down.

\subsection{Discussion}

Throughout this section, we have assumed only one term
of the form~(\ref{s}).
This cannot be justified by any symmetry arguments,
since the only global symmetry we have left is an $SU(n-2)$
global symmetry, which can be invoked to rule out terms 
such as $b^A f{\bar f}$ with $A=2\ldots n-1$.
However, our qualitative results remain unaffected
even if several terms of the form~(\ref{s}), with different 
composites appear, unless some special cancellation occurs.
First, we note that vevs of the baryons and trilinears 
differ by the ``small coupling'', either $\al$ or $\lam$,
which gives at least an order of magnitude difference.
Thus, if we use a baryon to generate the messenger masses,
through the term 
$b f \cdot{\bar f}/M^{n-2}$,
additional terms such as 
$Y f \cdot{\bar f}/M^2$,
are negligible, and vice versa.
Furthermore, in the examples we constructed with $S=b^n$,
$b^1$ had comparable, or smaller vevs. 
Its presence in the superpotential would thus not affect the 
results dramatically, 
unless its coupling to the messengers appears
with a different coefficient, such that some combination of expectation
values conspires to cancel.
The same is true for the trilinears.

Let us now summarize the different energy scales that appear in these models.
For concreteness, take the $SU(7)\times SU(6)$ model with $M = M_{GUT}$ 
and $S = b^7/M^3$. 
The $SU(6)$ group is broken at the scale $v \sim 10^{14}$~GeV, 
which is also the mass scale of the fields $Q$ and $L$.
$SU(7)$ then confines at the scale $\Lambda \sim 10^{10}$~GeV.

The light fields of the supersymmetry-breaking sector are  the baryons 
$b^1$, $b^7$,
whose scalar and fermion components have masses of $10^{4.5}$~GeV,
and the baryons $b^A$ with $A=2\ldots 6$, which make up one fundamental 
of the unbroken global $SU(5)$, whose fermion components are massless 
(as required by anomaly matching), and whose scalar components have masses
of order  $10^{4.5}$~GeV.
Finally, the messenger masses are also  of order  $10^{4.5}$~GeV,
and the supersymmetry-breaking scale is 
$\sqrt{F_0} \sim 10^9$~GeV.

New massless fermion species, beyond those present in the Standard Model,
may spoil the predictions of standard nucleosynthesis theory,
if they contribute significantly to the entropy at the time of 
nucleosynthesis ($T\sim 1$~MeV)~\cite{kolbturner}.
However, the massless fermions of our models--the fermion components 
of $b^{2\ldots (n-2)}$--interact extremely weakly, so that their decoupling 
temperature is very high.
Consequently, their contribution to the entropy at the time of nucleosynthesis
is negligible.
To see this, note that at sufficiently low temperatures,
the intreractions of these fermions are described by the low-energy Lagrangian
derived from~(\ref{spotential}), (\ref{kahler}) (see~\cite{wess}).    
Their dominant interaction comes from a 4-fermion term suppressed by 
$v^{-2}$.
The rate of this interaction is therefore $\Gamma \sim v^{-4} T^5$,
which is comparable to the expansion rate $H \sim T^2/M_{Planck}$
only for 
$T \geq 10^{13}$~GeV\footnote{
At this temperature the low-energy theory is
no longer valid for specific models. 
In all the examples we considered however, 
the low energy theory is valid below, say,  10$^9$~GeV, 
where the rate is even smaller.}.

As mentioned above, our models also contain exotic scalars
and fermions with masses around 10$^4-10^5$~GeV. 
These would be present in generic models of the type 
we consider, whereas the existence of the massless fermions
is a specific feature of the $SU(n)\times SU(n-1)$
supersymmetry-breaking sector.
The interactions of this exotic matter are again extremely weak.
The dominant fermion interaction is the 4-fermion interaction
mentioned above.
The scalar-interaction Lagrangian derived from~(\ref{spotential}), (\ref{kahler})
contains couplings involving only scalar baryons, 
as well as couplings of scalar baryons to scalar messengers.
(Note that all scalar baryons have couplings to the messengers through the
K\" ahler potential~(\ref{kahler})).  
These couplings are very small.
The typical 4-scalar term has a coefficient $(v/M)^{2n-4}$,
and higher order terms are further suppressed by negative powers
of $v$.
Therefore, the interactions of these fields are not thermalized at
 temperatures
for which the low-energy effective theory is valid.

In fact, the maximum reheating temperature after inflation
is constrained by requiring that the decay of the LSP to the gravitino
does not overclose the universe~\cite{hitoshigravitino}. 
In our case, the gravitino mass is ${\cal O}(1)$~GeV,
for which the authors of ref.~\cite{hitoshigravitino} 
conclude that the reheating temperature cannot exceed~${\cal O}(10^8)$~GeV.
Therefore, once diluted by inflation, 
the baryon fields are not produced 
thermally\footnote{
The baryons are reminiscent of the moduli of Hidden Sector models
in that they interact very weakly and have large 
vevs.
One may therefore worry about the analog of the Polonyi 
problem~\cite{banksnelson}.
However, here $v$ is at most $10^{14}$~GeV and the mass of the baryons is
$10^4-10^5$~GeV so the ratio of baryon density to entropy,
which scales like $v^2 m_b^{-1/2}$ is about nine orders of magnitude 
smaller than in the supergravity case.
}. 

Finally we note that the superpartner spectrum of our models
is identical to that of the models of~\cite{dnns},
since the masses of the Standard Model superpartners
only depend on the messenger masses.
The only different feature, from the point of view of 
phenomenology, is that the supersymmetry-breaking scale is relatively 
high,
$\sqrt{F_0} \sim 10^9$~GeV,
so that the decay of the LSP to the gravitino would not occur inside
the detector.
In this respect our models are similar to the models 
of~\cite{pt}, \cite{berkeley}.

\section{Conclusions}

In this paper we explore the possibility of eliminating
the fundamental singlet of existing models of gauge-mediated supersymmetry
breaking,
by introducing a non-renormalizable superpotential term 
that couples
the messengers to a chiral gauge-invariant of the 
supersymmetry-breaking theory.

We show
that to obtain viable models without 
${\cal O}(10^{-9})$ couplings,
the theory used as the
supersymmetry-breaking sector
should have 
a non-renormalizable superpotential.

We then construct several examples with 
non-renormalizable
$SU(n)\times SU(n-1)$
theories as the supersymmetry-breaking sector,
taking different gauge-invariants
to replace the fundamental singlet.
These examples only require couplings of order
$10^{-4}-10^{-3}$ for $M=M_{GUT}$, 
and of order $10^{-2}-10^{-1}$ for 
$M\leq 10^{15}$~GeV, where $M$ is the 
suppression scale of the non-renormalizable terms.

\section*{Acknowledgements}

I thank G.~Anderson, H.~C.~Cheng, 
J.~Lykken, Y.~Nir and
R.~Rattazzi for valuable discussions.
I am especially grateful to E.~Poppitz and S.~Trivedi
for many useful conversations and for their comments on
the manuscript.
The author
acknowledges the support of DOE contract DE-AC02-76CH0300.

\vspace{1.5 cm}
\noindent
Note: After this work was completed I became aware
of the work of Haba, Maru and Matsuoka~\cite{haba},
hep-ph/9703250 and 
hep-ph/9612468, where the 
same idea is considered, 
but with different supersymmetry-breaking theories. 
The model of hep-ph/9612468
uses the renormalizable 3-2 model 
with a coupling of order $10^{-9}$,
in agreement with the discussion of section~2.

\begin{appendix}
\section*{Appendix}
\renewcommand{\theequation}{A.\arabic{equation}}
\setcounter{equation}{0}

This appendix summarizes some details regarding the minimum we consider. 

It is convenient to work in terms of the baryon fields $b^A$.
The K\" ahler metric can be derived from~(\ref{kahler}),
\beq
\label{metric}
g_{AB} \ =\  k^{{1\over n-1}-2}\, 
\left(-{n-2\over n-1} b_B^\dagger b_A \, +\,
k\,\delta_{AB} \right) \ ,
\eeq
with  $k = b^\dagger_A b^A$, 
and can be easily inverted to get,
\beq
\label{invmetric}
g_{AB}^{-1} \ =\ k^{-{1\over n-1}}\, 
\left( (n-2)\, b_A^\dagger b_B \, +\,
k\,\delta_{AB} \right) \ .
\eeq
The potential is then given by,
\beq
g^{-1}_{AB}\, W_A\, W_B \ ,
\eeq
with, using~(\ref{spotential}),
\beqa
W_1 &=& \al \ , \\
W_n &=& {1\over n}\,
\left( \lambda^{n-1}\, \Lambda^{2n+1}\right)^{{1\over n}}\,
(b^n)^{{1\over n}-1} \, +\, \beta \ , \\
W_A &=& 0 \ \ \ \  for \ A = 2\ldots n-1\ . 
\eeqa

One can then show analytically that the potential is minimized for
$b^1 = r b^n \equiv r v^{n-1}$,
where $0 < r < \sqrt{n-1}/2$ 
is determined from
\beq
\label{betoveral}
{\beta\over\alpha}\ =\ 
{
(r^2+1)\,(n\,r^2-r^2+2)\, P -3\, r^4 +
8n\, r^4-2n + 5r^2
+n^2r^2-4nr^2-3 n^2 r^4+2 
\over
2(n-1)\, (2n-2+r^2)\, r^3 } \ ,
\eeq
where
\beq
P\ = \  \sqrt{n-1}\, \sqrt{n-1-4r^2} \ ,
\eeq  
and with $v$ given by,
\beq
{\Lambda\over  v} \ =\
\left(n q^{-1} \al \lam^{-{n-1\over n }} 
\vom^{n-4} \right)^{{n\over 2n+1}} \ ,
\eeq
where
\beq
q\ = \ -{(n-1)\over n\, (n-2)}
\, {r\, (P + n -r^2 -1)
\over
 (r^2 + 1) } \ .
\eeq

Note that $q < 0$. We therefore take $\al$ to be negative.
The bounds on $\al$ appearing in the text refer to its absolute
value.

The functions $F_1$, $F_n$ defined in~(\ref{barf})
are then given by,
\beqa
F_1\ &=&\
1\ + (n-1)\, r^2 \, 
+\, (n-2)\, r \left( {1\over q} \, +\, {\beta\over\al}\right) \ , \\
F_n\ &=&\
(n-2)\, r \, +\, (n-1+r^2)\, \left( {1\over q} \, 
+\, {\beta\over\al}\right) \ .
\eeqa

The simplest flat direction that results in the 
baryon configuration~(\ref{b1n})
is of the form
$R_{Ai} = a v \delta_{Ai}$ for $A=2\ldots n-1$, 
$R_{11} = a^{-(n-2)} v$, and
$R_{n1} = r a^{-(n-2)} v$,
where $a= (1+r^2)^{{1\over 2(n-1)}}$,
and the second index on $R$ is the $SU(n-1)$ gauge index.

We then have
\beqa
f_{R_1}\ &=&\
{1\over n-1}\, (1+r^2)^{-{n-2\over 2(n-1)}}\,
\left(
1\, -\, {(n-2)\, F_1\over r\,F_n}\right)\, F_n \ ,\\
f_{R_n}\ &=&\
-\, {n-2\over n-1}\, (1+r^2)^{-{n-2\over 2(n-1)}}\, r\,
\left(
1\, -\, { F_1\over (n-2)\, r\,F_n}\right)\, F_n \ ,\\
f_{R_A}\ &=&\
{1\over n-1}\, (1+r^2)^{1\over 2(n-1)}\,
\left(
1\, +\, {F_1\over r\,F_n}\right)\, F_n \ \ \
for  \ A=2\ldots n-1\ . 
\eeqa 

Finally, to obtain the expectation values of the trilinears 
$Y_{IA}$, recall that
$Y_{IA} = L_I \cdot Q \cdot R_A$,
and that $L$ and $Q$ are the heavy flavors  of $SU(n)$
with a mass matrix $m = diag(R_{11},..,R_{(n-1)(n-1)})$.
Therefore, using~\cite{seibergexact}
\beq
\langle Q\cdot L \rangle \ =\ \left( \Lambda^{2n+1}
\det{m} \right)^{{1\over n}}\,
m^{-1} \ , 
\eeq
one finds~(\ref{yii}).

\end{appendix}

\nc{\ib}[3]{ {\em ibid. }{\bf #1} (19#2) #3}
\nc{\np}[3]{ {\em Nucl.\ Phys. }{\bf #1} (19#2) #3}
\nc{\pl}[3]{ {\em Phys.\ Lett. }{\bf #1} (19#2) #3}
\nc{\pr}[3]{ {\em Phys.\ Rev. }{\bf #1} (19#2) #3}
\nc{\prep}[3]{ {\em Phys.\ Rep. }{\bf #1} (19#2) #3}
\nc{\prl}[3]{ {\em Phys.\ Rev.\ Lett. }{\bf #1} (19#2) #3}

\end{document}